\documentclass[%
 aip,
 amsmath,amssymb,
 reprint
]{revtex4-1}

\usepackage{graphicx}
\usepackage{dcolumn}
\usepackage{bm}

\usepackage[utf8]{inputenc}
\usepackage[T1]{fontenc}
\usepackage{mathptmx}
\usepackage{etoolbox}
\usepackage{lineno}
\usepackage[dvipsnames]{xcolor}

\newcommand{\upd}{\mathrm{d}}

\makeatletter
\def\@email#1#2{%
 \endgroup
 \patchcmd{\titleblock@produce}
  {\frontmatter@RRAPformat}
  {\frontmatter@RRAPformat{\produce@RRAP{*#1\href{mailto:#2}{#2}}}\frontmatter@RRAPformat}
  {}{}
}%
\makeatother
\begin{document}

\preprint{AIP/123-QED}

\title
{
Microrheology with rotational Brownian motion
}
\author{Yasuya Nakayama}
 \email{nakayama@chem-eng.kyushu-u.ac.jp}
\affiliation{ 
Department of Chemical Engineering,
Kyushu University,
Nishi-ku,
Fukuoka 819-0395,
Japan
}%

\date{\today}

\begin{abstract}
Passive rotational microrheology (RMR) for evaluating the dynamic modulus \(G^*\) of a suspending fluid through the rotational Brownian motion of a spherical probe particle is validated using direct numerical simulations (DNS) of Brownian motion in a viscoelastic fluid. 
Two methods of RMR  are compared: an inertialess RMR based on the Generalized Stokes--Einstein relation for rotational diffusion (RGSER) and the full RMR  based on the generalized Langevin equation for rotation, which accounts for fluid and particle inertia. 
Our analysis, performed using DNS of the fluctuating Oldroyd-B fluid, reveals that 
inertialess RMR accurately estimates \(G^*\) for \(\omega\lambda \alt 1\) but deviates significantly at high frequencies. In contrast, the full RMR improves \(G^*\) estimation accuracy up to the frequency \(\omega \approx \tau_{s}^{-1}=\eta_{s}/\rho_{f}a^{2}\), where fluid inertia becomes relevant. However, in the ballistic regime (\(t \ll \tau_{s}\)), particle inertia dominates, making accurate \(G^*\) evaluation challenging even with the full RMR. This study clarifies the applicability range of RMR. Additionally, rotational Brownian motion is turned out to be insensitive to periodic boundary conditions, which allows direct application to various mesoscale molecular simulations, including coarse-grained molecular dynamics, dissipative particle dynamics, and fluid dynamics simulations. In conclusion, rotational microrheology offers a promising approach for detailed rheological analysis in complex systems and conditions.
\end{abstract}

\maketitle

\section{Introduction}
Microrheology is a technique for measuring the rheological response of complex fluids at the microscale~\cite{Mason1995,Mason1997}. 
Colloidal beads are used as probe particles and their motion is tracked to evaluate rheology of a suspending medium. 
Specifically, the technique that utilizes Brownian motion driven by thermal fluctuations is called passive microrheology, while the technique that controls particle motion by applying an external force is called active microrheology~\cite{kimura2009microrheology,squires2010fluid,waigh2016advances,zia2018active,liu2018rheological}.
Compared with conventional macrorheometers, microrheology allows rheological measurements with small sample volumes, which is advantageous for applications dealing with soft, delicate, or inhomogeneous materials. 
Microrheology technique has been reported to be effective in evaluating the heterogeneity of complex fluids, such as food and gels~\cite{yang2017application}, the rheological response of cultured cells~\cite{hoffman2006consensus}, the rheology of collagen gels containing cells~\cite{cai2021dynamic}, and the internal rheology of living cells~\cite{yamada2000mechanics,xia2018microrheology}.

In addition to analyzing the translational motion of the probe particles, rotational microrheology~(RMR), which utilizes rotational motion, has also been studied.
Corresponding to the generalized Stokes--Einstein relation~(GSER) 
for translational Brownian motion, the rotational form of GSER 
for rotational Brownian motion (RGSER) has been 
introduced~\cite{cheng2003rotational}.
Furthermore, the applicability range of RGSER-based RMR has been theoretically examined based on the two-fluid model~\cite{schmiedeberg2005one}.
Experimentally, birefringent spherical particles have been synthesized, enabling passive RMR measurements using RGSER~\cite{andablo2005microrheology}, as well as measurements of hydrodynamic interactions and particle friction in dense colloidal suspensions~\cite{yanagishima2021particle}.
Additionally, active viscosity measurements using optical tweezers to control rotation have also been reported~\cite{bishop2004optical}.

The RGSER used in RMR does not consider the inertia of the fluid and particles, so the availability of RMR at high frequencies has not been clearly established.
Therefore, we refer to the RGSER-based RMR as the inertialess RMR.
By applying the generalized Langevin equation (GLE) to rotational motion, it is possible to consider 
full RMR taking inertia into account, but its effectiveness has not been fully explored.
Meanwhile, we have developed a direct numerical simulation~(DNS) of the Brownian motion of colloidal particles in a viscoelastic fluid driven by thermal fluctuations~\cite{nakayama2024simulating}.
This DNS allows the analysis of the Brownian motion of particles suspended in fluids with well-defined rheology.
Using this DNS, we have accounted for inertial effects while revealing finite system size effects in translational passive microrheology~\cite{nakayama2024simulating,nakayama2025brownian}.
In this study, using the rotational Brownian motion results of DNS, we verify the applicability of 
two methods of passive RMR: inertialess RMR and full RMR.
The paper is organized as follows.
In Sec.~\ref{sec:rgle_mr}, we outline the derivation of the rotational velocity autocorrelation function~(RVACF), rotational mean square displacement~(RMSD), and RMR relations based on the GLE for the rotational Brownian motion of a spherical particle and RGSER.
In Sec.~\ref{sec:dns_sspm}, we provide an overview of direct numerical simulations of spherical beads in viscoelastic flows driven by thermal fluctuations.
In Sec.~\ref{sec:results}, we discuss the results of the direct numerical simulation of the RMSD of 
a spherical bead and the estimation of the dynamic modulus using the RMR methods.
In Sec.~\ref{sec:conc}, the results are summarized, and the concluding remarks are made.

\section{\label{sec:rgle_mr}
Rotational diffusion microrheology}
This section provides an overview of the relationship between the RVACF, 
the RMSD of a spherical particle, and the dynamic modulus of the suspending fluid based on the GLE of angular velocity of the particle.
\subsection{Generalized Langevin equation of a rotating particle}
Consider an isolated spherical particle of mass \(M\) and radius \(a\) immersed in a fluid of density \(\rho_f\). The angular velocity of the rotation of the particle \(\bm{\Omega}\) follows the following GLE: 
\begin{linenomath}
\begin{align}
\label{eq:gle}
 I_{p}\frac{\upd\bm{\Omega}}{\upd t}
&=-\int_{-\infty}^{t}\upd s\zeta_r(t-s)\bm{\Omega}(s)+\bm{N}_{R}(t),
\end{align}
\end{linenomath}
where \(I_{p}=2M a^2/5\) is the moment of inertia of the Brownian particle.
The first term on the right-hand side of Eq.~(\ref{eq:gle}) represents the hydrodynamic drag where the time-dependent rotational friction coefficient \(\zeta_r(t)\) is determined by hydrodynamic interactions and the rheology of the fluid.
\(\bm{N}_{R}\) represents the random torque caused by thermal fluctuations in the fluid and is independent of the particle's angular velocity history, {\itshape i.e.}, \(\langle \bm{N}_{R}(t>0)\bm{\Omega}(0)\rangle = \bm{0}\), where \(\left\langle \cdot\right\rangle\) denotes the statistical average. 
The variance of \(\bm{N}_{R}\) is determined by the friction coefficient according to the fluctuation-dissipation theorem~\cite{Landau1980} as
\(
\langle N_{R,i}(t)
N_{R,j}(0)\rangle
=\delta_{ij}k_{B}T\zeta_{r}(t),
\)
where \(\delta_{ij}\) is the Kronecker delta, \(T\) is the temperature, and \(k_{B}\) is the Boltzmann constant.

The equation governing RVACF is obtained by multiplying the GLE~[Eq.~(\ref{eq:gle})] by \(\bm{\Omega}(0)\) and taking the statistical average.
Due to the isotropy of the particle, each component of the angular velocity vector is independent, so it is sufficient to consider the RVACF in one direction, \(C_{R}(t)=\langle\Omega_x(t)\Omega_x(0)\rangle\). 
Therefore, the equation that \(C_{R}\) follows is: 
\begin{linenomath}
\begin{align}
\label{eq:rvacf_evolution}
 I_{p}\frac{\upd C_{R}}{\upd t}
&=
-\int_{0}^{t}\upd s \zeta_{r}(t-s)C_{R}(s),
\end{align}
\end{linenomath}
By taking the Laplace transform 
\(
 \widehat{f}(\omega) 
  = \left[\int_{0}^{\infty}\upd t f(t)\exp\left(-zt\right)\right]_{z=i\omega},
\) of Eq.~(\ref{eq:rvacf_evolution}), we obtain
\begin{linenomath}
\begin{align}
\label{eq:rvacf_omega}
 \widehat{C_{R}}(\omega) 
&=
\frac{k_{B}T}{
i\omega I_{p}+\zeta_{r}(\omega)
},
\end{align}
\end{linenomath}
where \(C_{R}(t=0)=\langle \Omega_{x}^{2}\rangle = k_{B}T/I_{p}\) 
is used.
The frequency-dependent rotational friction coefficient, 
\(\zeta_{r}(\omega)\), can be derived by solving the Stokes 
equation for the flow around a single isolated sphere oscillatory 
rotating with angular frequency \(\omega\) and evaluating the 
resulting 
torque~\cite{lamb_hydrodynamics6th,Landau1959fluid,zatovsky1969,berne1972hydrodynamic,Lisy2004}, which is represented as follows:
\begin{linenomath}
\begin{align}
\label{eq:r-friction}
 \zeta_{r}(\omega)
&=8\pi a^{3}\eta^{*}(\omega)\left(1+\frac{i\omega 
 \rho_{f}a^{2}/3\eta^{*}(\omega)}{1+\sqrt{i\omega \rho_{f}a^2/\eta^{*}(\omega)}}\right),
\end{align}
\end{linenomath}
where the complex viscosity \(\eta^{*}(\omega)\) is defined from the dynamics modulus \(G^{*}(\omega)=G'(\omega)+iG''(\omega)\) as \(\eta^{*}(\omega)=G^{*}(\omega)/i\omega\).
Equation~(\ref{eq:rvacf_omega}) relates the rheology of the suspending fluid to the RVACF of the Brownian particle.

\subsection{Rotational microrheology relation}
To analyze the change of orientation of a Brownian particle, a unit orientation vector \(\bm{p}\) fixed on the particle is considered.
The time evolution of \(\bm{p}\) follows 
\(
 \frac{\upd \bm{p}}{\upd t}  = \bm{\Omega}\times \bm{p}
\).
Following Refs.~\onlinecite{kammerer1997dynamics,Mazza2007,Hunter2011}, 
the vector rotational displacement is defined as follows:
\begin{linenomath}
\begin{align}
\label{eq:vector_rotational_displacement}
 \Delta\bm{\phi}(t) &= \int_{0}^{t}\bm{\omega}(s)\upd s,
\end{align}
\end{linenomath}
where \(\bm{\omega}=\bm{p}\times\upd\bm{p}/\upd 
t=\bm{\Omega}-(\bm{p}\cdot\bm{\Omega})\bm{p}\).
Note that since \(\bm{p}\) moves on the surface of a unit sphere, both \(\bm{p}\) and \(\bm{\omega}\) have two degrees of freedom, 
respectively, meaning that \(\langle\bm{\omega}^{2}\rangle = 
\langle\bm{\Omega}^{2}\rangle-\langle 
\left(\bm{p}\cdot\bm{\Omega}\right)^{2}\rangle = 2k_{B}T/I_{p}\).
RMSD is defined as 
\(\langle\Delta\bm{\phi}^{2}(t)\rangle\).
The definition of \(\Delta \bm{\phi}\)  Eq.~(\ref{eq:vector_rotational_displacement})
leads to 
\begin{linenomath}
\begin{align}
\label{eq:rmsd}
 \langle\Delta\bm{\phi}^{2}(t)\rangle
&=2\int_{0}^{t}\left(t-s\right)\langle\bm{\omega}(s)\cdot\bm{\omega}(0)\rangle \upd s.
\end{align}
\end{linenomath}
By assuming a relation 
\(\langle\bm{\omega}(t)\cdot\bm{\omega}(0)\rangle\approx 
2C_{R}(t)\), the RMSD can be evaluated with the RVACF.

RMSD asymptotically approaches \(4D_{r}t\) for longer times than the angular correlation time. 
According to the Stokes--Einstein--Debye relation, the rotational diffusion coefficient is \(D_{r}=k_{B}T/(8\pi\eta_{0}a^{3})\), where \(\eta_{0}\) is the zero-shear-rate viscosity.
On the other hand, in the short-time region where particle inertia dominates, \(t \ll I_{p}/(8\pi \eta_{s}a^{3})\), ballistic motion is expected and is described by \(\left\langle \Delta\bm{\phi}^{2}(t)\right\rangle \approx 2\left(k_{B}T/I_{p}\right)t^{2}\), where \(\eta_{s}\) is the viscosity of the solvent.

From Eq.~(\ref{eq:rvacf_omega}), when the inertia of both the particle and the fluid can be neglected (\(I_{p}, \rho_{f} \to 0\)), we obtain the generalized Stokes--Einstein relation for rotation~(RGSER)~\cite{cheng2003rotational}:
\begin{linenomath}
    \begin{align}
    \label{eq:rgser}
 \eta_{RGSER}^{*}(\omega) &= \frac{k_{B}T}{8\pi a^{3}\widehat{C_{R}}(\omega)}.
    \end{align}
\end{linenomath}
This RGSER, similar to GSER
for translational motion, is expected to hold at sufficiently low frequencies~\cite{karim2012determination,nakayama2024simulating}.
In the frequency domain where hydrodynamic interactions and fluid viscoelasticity dominate, \(\eta^{*}(\omega)\) can be estimated using the following relation: 
\begin{linenomath}
    \begin{align}
    \label{eq:rgle_mr}
\eta^{*}
\left(1+\frac{1}{3}\frac{i\omega\rho_{f}a^{2}/\eta^{*}}{1+\sqrt{i\omega\rho_{f}a^{2}/\eta^{*}}}\right)
&=
\eta^{*}_{RGSER}-\frac{i\omega I_{p}}{8\pi a^{3}},
    \end{align}
\end{linenomath}
derived from Eqs.~(\ref{eq:rvacf_omega}) and (\ref{eq:r-friction}).
Equation~(\ref{eq:rgle_mr}) can be solved using iterative methods.
\section{\label{sec:dns_sspm}
Direct simulation of rotational Brownian motion}
To investigate the validity of the RMR
analysis, we perform direct numerical simulations of the rotational diffusion of a Brownian particle in a fluid with specified viscoelastic properties.
In this section, we provide an overview of the stochastic Smoothed Profile method~(sSPm)~\cite{nakayama2024simulating}, which is a direct simulation method to solve the coupled motion of particles and thermally driven fluid based on the fluctuating hydrodynamics~\cite{Landau1980} and fluctuating viscoelasticity~(FVE) frameworks~\cite{hutter2020fluctuating,Hutter2018,Hutter2019,Hutter2020}.
\subsection{Fluctuating Oldroyd-B fluid}
Based on the fluctuation--dissipation theorem, several continuum fluid simulations have been proposed to solve viscoelastic flows driven by thermal fluctuations at small scales~\cite{Voulgarakis2009,Voulgarakis2010,Vazquez-Quesada2009,Vazquez-Quesada2009a,Vazquez-Quesada2012,Hutter2018,Hutter2019,Hutter2020,Hohenegger2017,Paul2018,Paul2019}. 
In this study, we consider the fluctuating Oldroyd-B model based on FVE framework~\cite{Hutter2018,Hutter2019,Hutter2020,hutter2020fluctuating}.
To represent the elastic degrees of freedom, we use the conformation tensor \( \bm{C} \), a second-order tensor. This tensor \( \mathbf{C} \) characterizes the elastic degrees of freedom and gives rise to elastic stress depending on the deviation from the equilibrium state.
The conformation tensor is microscopically derived from the ensemble average of the dyad of end-to-end vectors in the dumbbell model~\cite{bird1987dynamics}.
The elastic stress \(\bm{\sigma}_{p}\) 
is expressed as a function of \(\bm{C}\) as \(\bm{\sigma}_{p} = \frac{\eta_{p}}{\lambda}\left(\bm{C}-\nu_{p}\bm{I}\right)\),
where \(\lambda\), \(\eta_p\), and \(\bm{I}\) are the relaxation 
time, the polymer viscosity, and unit tensor, respectively.
\(\nu_{p}=1-4k_{B}T\lambda/\eta_{p}\Delta^3\) is a constant factor that determines the polymer conformation free energy minimizer due to the finite number of dumbbells correction~\cite{hutter2020fluctuating}, 
and \(\Delta^{3}\)
is the representative volume of coarse-grained stochastic description and is associated with a single node in discretized space.

The conformation tensor field \(\bm{C}\) evolves according to the flow and thermal fluctuations as
\begin{linenomath}
\begin{widetext}
\begin{align}
\label{eq:fluct_c_oldroyd_eq}
\upd\bm{C}
&=
\left(-\bm{u}\cdot\nabla\bm{C}+(\nabla\bm{u})^{t}\cdot\bm{C}+\bm
{C}\cdot(\nabla\bm{u})-\frac{1}{\lambda}\left(\bm{C}-\bm{I}\right)
\right)\upd t+
\left(\bm{b}\cdot \upd\bm{W}+\upd\bm{W}^{t}\cdot\bm{b}^{t}\right),
\end{align}
\end{widetext}
\end{linenomath}
where \(\bm{u}\) is the velocity field,
\(\bm{b}\) is a second-order tensor that satisfies
\(\bm{C}=\bm{b}\cdot\bm{b}^t\), and \((.)^t\) indicates a transpose.
The second-order tensor \( \upd\bm{W} \) represents a random force increment that satisfies the following fluctuation--dissipation relation: 
\begin{linenomath}
\begin{align}
\left\langle \upd W_{ij}(\bm{x},t)\right\rangle  & = 0,
\label{eq:random_mean}
\\
\left\langle \upd W_{ij}(\bm{x},t)\upd W_{kl}(\bm{x}',t')\right\rangle 
& = 
\frac{k_BT}{\eta_{p}\Delta^{3}}\left(1-\phi\right)
\delta_{ik}\delta_{jl}
\delta_{\bm{x},\bm{x}'}
\delta_{t,t'}
\upd t, \label{eq:random_sigma}
\end{align}
\end{linenomath}
where \(\delta(\bm{x}-\bm{x}')\approx\delta_{\bm{x},\bm{x}'}/\Delta^3\).
\(\phi(\bm{x},t)\) is the domain indicator function, which is unity inside the particle and zero outside the particle.

Although \(\bm{C}\) is physically positive definite, the random driving by the stochastic partial differential equation~(SPDE)~(\ref{eq:fluct_c_oldroyd_eq}) does not guarantee the positive definiteness of \(\bm{C}\).
To solve this problem, the time evolution of the \(\bm{b}\) tensor rather than the \(\bm{C}\) tensor itself is solved, and then \(\bm{C}\) is constructed from \(\bm{b}\)~\cite{Hutter2018,Hutter2019,Hutter2020,hutter2020fluctuating,nakayama2024simulating}.
Further details on the \(\bm{b}\)-formulation of Eq.~(\ref{eq:fluct_c_oldroyd_eq}) are described in Ref.~\onlinecite{nakayama2024simulating}.

\subsection{Stochastic Smoothed Profile method}
To solve the coupled viscoelastic flow and particle motion, SPm~\cite{Nakayama2008,yamamoto2021smoothed} is used.
As the flow is driven by thermal fluctuations, 
SPm is applied to SPDE based on fluctuating hydrodynamics~\cite{nakayama2024simulating}.

Consider a spherical particle suspended in a thermally fluctuating viscoelastic fluid. The motion of the particle is described by 
\(\dot{\bm{R}}=\bm{V}\), 
\(M\dot{\bm{V}}=\bm{F}^H\),
and 
\(I_{p}\dot{\bm{\Omega}}=\bm{N}^H\),
where $\bm{R}$, and $\bm{V}$ are the position and velocity of the particle, respectively;
$\bm{F}_i^H$ and $\bm{N}_i^H$ are the hydrodynamic force and
torque of the fluid~\citep{Nakayama2008,Molina2016}, 
respectively, which are responsible for the hydrodynamic drag and fluctuating stress.

In SPm, the velocity field is defined over the whole domain to include 
both the matrix fluid and the particles as 
\(\bm{u}(\bm{x},t) = \left(1-\phi\right)\bm{u}_{f}
+\phi \bm{u}_{p}\),
where \(\bm{u}_{f}\) represents the velocity in the matrix fluid 
domain; the particle velocity field $\phi\bm{u}_{p}$ is 
constructed with the positions and velocities of the particles as
\(\phi\bm{u}_{p}
=
\phi
\left[
\bm{V}+\bm{\Omega}\times\left(\bm{x}-\bm{R}\right)
\right]\).
In SPm, the interface profile of \(\phi\) does not exhibit a step function. Instead, a diffuse interface domain characterized by a scale \(\xi\) is explicitly introduced, allowing \(\phi\) to be tracked on a fixed spatial grid.
Details of the specific definition of \(\phi\) are reported in Ref.~\onlinecite{Nakayama2005,yamamoto2021smoothed}.
The velocity field is governed by the Navier--Stokes equation, 
with thermally fluctuating stress in the form of a SPDE, and the mass continuity equation as follows: 
\begin{linenomath}
\begin{align}
\rho_{f}\upd\bm{u} &= 
\left[-\rho_{f}\bm{u}\cdot\nabla\bm{u} 
-\nabla p +\nabla\cdot \left(2\eta_s\bm{D} +\bm{\sigma}_{p}\right)\right]\upd t 
\notag
\\
&
+\nabla\cdot\upd \bm{\sigma}^{r}
+\rho_{f}\phi\bm{f}_{p}\upd t,
\label{eq:fluctuating_NS}
\\
\nabla\cdot\bm{u} & =  0,
\label{eq:incomprssibility}
\end{align}
\end{linenomath}
where \(p\) and
\(\bm{D}=\left(\nabla\bm{u}+\nabla\bm{u}^{t}\right)/2\) are the pressure and
the strain-rate tensor, respectively.
The body force \(\phi\bm{f}_{p}\) is a penalty force that is used to enforce the rigid-body constraints within the particle domain; the 
implementation of this force is detailed in Ref.~\onlinecite{nakayama2024simulating}.
The fluctuating stress increment $\upd\bm{\sigma}^{r}$ is a 
spatio-temporally white random process whose variance is 
determined by the fluctuation--dissipation relation as follows~\cite{Landau1980}: 
\begin{linenomath}
\begin{align}
\left\langle \upd\sigma_{ij}^{r}(\bm{r},t)\right\rangle  & = 0,
\\
\left\langle \upd\sigma_{ij}^{r}(\bm{r},t)\upd\sigma_{k l}^{r}(\bm{r}',t')
\right\rangle  
& = \frac{2\eta_{s}k_{B}T(1-\phi)}{\Delta^{3}}
(\delta_{ik}\delta_{jl}+\delta_{jk}\delta_{il})
\delta_{\bm{r},\bm{r}'}\delta_{t,t'}\upd t,
\label{eq:random_stress_newton}
\end{align}
\end{linenomath}
The non-slip boundary condition for the velocity field is assigned at particle surfaces. 
The forces on the particle \(\bm{F}^{H}\) and \(\bm{N}^{H}\) due to fluid flow are evaluated based on the momentum exchange between the particle and fluid~\cite{nakayama2024simulating}.

\subsection{Simulation conditions}
The complex viscosity of the Oldroyd-B model is
\begin{align}
\label{eq:ob_eta}
 \eta^{*}(\omega) &= \eta_{s}+\frac{\eta_{p}}{1+i\omega \lambda},
\end{align}
from which, the zero-shear-rate viscosity is identified as \(\eta_{0}=\eta_{s}+\eta_{p}\), and the viscosity ratio \(\beta=\eta_{s}/\eta_{0}\) is defined.
Therefore, the dynamic modulus of the Oldroyd-B model is characterized by \(\eta_{0}, \beta\), and \(\omega\lambda\).
We perform DNS of a single Brownian particle to calculate RVACF and RMSD~\cite{nakayama2024simulating}.
The simulation is carried out under periodic boundary conditions with a linear domain size of \(L=64\Delta\). The system domain was uniformly discretized with a spacing of \(\Delta\), which is the unit length of the system.
The particle radius was set to \(a=5\Delta\), so the particle volume fraction is approximately 0.002.
The interface scale of the smoothed profile is set to \(\xi=2\Delta\).
Two cases of the Oldryod-B model are considered, with the parameters shown in the Table~\ref{tab:parameters}.
Further details on the DNS are described in Ref.~\onlinecite{nakayama2024simulating}.

\begin{table}[]
    \centering
    \caption{\label{tab:parameters}
    Input parameters for the Oldroyd-B model. The simulation time unit is \(\tau_{0}=\Delta^{2}\rho_{f}/\eta_{s}\).}
\begin{ruledtabular}
    \begin{tabular}{cccc}
             \(\beta\) & \(\eta_{p}/\eta_{s}\) & \(\eta_{0}/\eta_{s}\) & \(\lambda/\tau_{0}\) \\
         \hline
         0.1 & 9 & 10 & 1000 
         \\
         0.5 & 1 & 2 & 1000 
         \\
    \end{tabular}
\end{ruledtabular}
\end{table}
\section{\label{sec:results}
Results and Discussion}
\subsection{RVACF and RMSD}
\begin{figure}
    \centering
    \includegraphics[width=246.0pt]{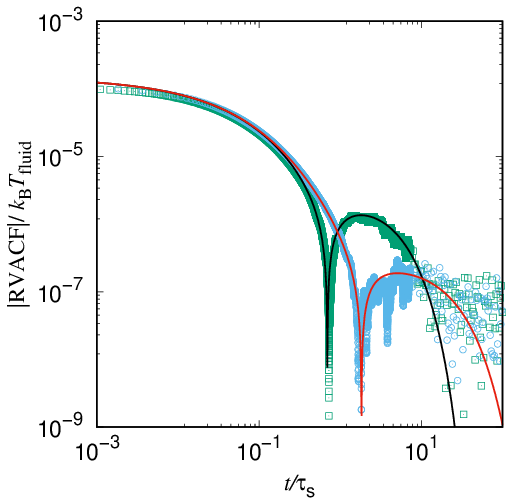}
    \caption{
    \label{fig:rvacf_ob2}
    Rotational velocity autocorrelation function~(RVACF)
    of a particle suspended in 
 Oldroyd-B fluids with \(\beta=0.1\) and 0.5 at a volume fraction of 0.002:
 The DNS results are plotted as green open squares for \(\beta=0.1\) and blue open circles for \(\beta=0.5\).
 The lines are the predicted values of the generalized Langevin equation~(GLE): black line for \(\beta=0.1\)) and red line for \(\beta=0.5\).
    }
\end{figure}

\begin{figure}
    \centering
    \includegraphics[width=246.0pt]{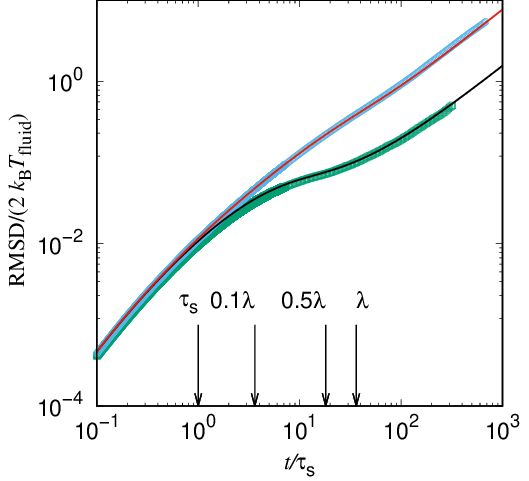}
    \caption{\label{fig:rmsd_ob2}
Rotational mean-square displacement~(RMSD) of a particle suspended in 
 Oldroyd-B fluids with \(\beta=0.1\) and 0.5 at a volume fraction of 0.002:
 The DNS results are plotted as green open squares for \(\beta=0.1\) and blue open circles bfor \(\beta=0.5\).
 The lines are the predicted values of the generalized Langevin equation~(GLE): black line for \(\beta=0.1\)) and red line for \(\beta=0.5\).
} 
\end{figure}

The RVACF and RMSD of a Brownian particle in the Oldroyd-B fluid for the cases of \(\beta=0.1\) and 0.5 are shown in Figs.~\ref{fig:rvacf_ob2} and \ref{fig:rmsd_ob2}, respectively. 
The RVACF (Fig.~\ref{fig:rvacf_ob2}) does not exhibit a monotonically decreasing behavior as observed in the Newtonian fluid ~\cite{hocquart1983long}. Instead, it oscillates and shows anticorrelation for \(t >\tau_{s} = \rho_{f}a^{2}/\eta_{s}\). 
This is due to the memory effect of fluid flow induced by the past motion of a particle on the subsequent dynamics. In Newtonian fluids, this effect leads velocity correlation to persist, specifically the power-law decay of RVACF\(\sim t^{-5/2}\).~\cite{zatovsky1969,berne1972hydrodynamic,Lisy2004}
On the other hand, in viscoelastic fluids, the presence of unrelaxed elastic degrees of freedom has an additional effect, leading to a reversal of angular velocities in the RVACF.

The RMSD (Fig.~\ref{fig:rmsd_ob2}) shows a similar behavior to the translational MSD~\cite{nakayama2024simulating}.
Namely, for \(t \ll \tau_{s}\), ballistic behavior is observed, while for \(\beta\lambda < t < \lambda\), subdiffusive behavior is found. For \(t \gg \lambda\), normal diffusion is evident.
The subdiffusion regime corresponds to the anticorrelation regime observed in RVACF and is dominated by the memory effects from both hydrodynamic interactions and viscoelasticity.
The rotational diffusion coefficient, evaluated in the normal diffusion regime at \(t \gg \lambda\), is \(D_{r} \approx 1.43 \times 10^{-8}\tau_{0}^{-1}\) when \(\beta = 0.5\) and \(D_{r} \approx 2.89 \times 10^{-9}\tau_{0}^{-1}\) when \(\beta = 0.1\).
This is consistent with the values obtained from the Stokes--Einstein--Debye relation as follows: 
\begin{align}
    D_{r} &=\frac{k_{B}T_{\text{fluid}}}{8\pi\eta_{0} a_{\text{eff}}^{3}},
\end{align}
where 
\(k_{B}T_{\text{fluid}}=\rho_{f}\langle \bm{u}^{2}\rangle\Delta^{3}/2\) is the temperature of the fluid, defined by the fluid velocity fluctuations, which in the numerical simulation was \(k_{B}T_{\text{fluid}}=0.99k_{B}T\) consistent with the set \(k_{B}T\) within 1 per cent; \(a_{\text{eff}}=1.052a\) is determined from the particle temperature~\cite{nakayama2024simulating}; these are effective values realized in DNS.
It should be noted that the system-size effect due to the PBCs, which occur in translational diffusion~\cite{dunweg1993molecular,Yeh2004,heyes2007self,heyes2007system}, does not affect the rotational diffusion coefficient.
This result can be understood as follows: the rotational motion is generated by the shear forces between the particle and the fluid. In other words, the longitudinal sound waves in the fluid, which affect the translational motion of the particle, are decoupled from the rotational diffusion. Therefore, the image cell particles associated with the PBCs do not weaken the rotational diffusion.
The solutions obtained from the rotational GLE [Eq.~(\ref{eq:rvacf_omega})] with the complex viscosity [Eq.~(\ref{eq:ob_eta})] without considering the PBCs are also plotted in Figs.~\ref{fig:rvacf_ob2} and \ref{fig:rmsd_ob2} and show good agreement with the simulated RVACF and RMSD.

\subsection{Rotational microrheology modulus
}
To perform the inertialess RMR
using RGSER~[Eq.~(\ref{eq:rgser})] or the full RMR based on GLE for rotation~[Eq.~(\ref{eq:rgle_mr})], it is necessary to calculate the Laplace transform of RMSD~\cite{Mason1997}.
The problem is that the measured RMSD is subject to statistical errors and because the RMR uses the second derivative of the RMSD, it is susceptible to rapid fluctuations caused by these statistical errors.
In this study, instead of performing the Laplace transform numerically on the RMSD data, we employ a method that fits the RMSD to a smooth function~\cite{karim2012determination,nakayama2024simulating}.
We use a fitting function introduced in Ref.~\onlinecite{karim2012determination} as
\begin{linenomath}
\begin{align}
\label{eq:msd_fitting}
\langle
\Delta\bm{\phi}^{2}(t)
\rangle
&=4D_{r}\lambda f_{LE}(t/\lambda) 
+\int_{0}^{\infty}\frac{\upd \tau}{\tau}h(\tau)f_{mKV}(t/\tau),
\end{align}
\end{linenomath}
Here, \(f_{\text{LE}}(t/\lambda) = 
\frac{t}{\lambda}+\left(e^{-t/\lambda}-1\right)\) and
\(f_{\text{mKV}}(t/\tau) = 1- \left(1+\frac{t}{\tau}\right)e^{-t/\tau}\),
with a piecewise power-law spectrum
\(h(\tau) =\sum_{j=1}^{n}g_{j}\tau^{\alpha_{j}}H(\tau_{j}-\tau)H(\tau-\tau_{j-1})\)
where \(H(.)\) is the Heaviside step function~\cite{baumgaertel1990relaxation}.
The details of the fitting model, including its Laplace transform, are
explained in Ref.~\onlinecite{nakayama2024simulating}.
We briefly describe the basic properties of this fitting functions in Eq.~(\ref{eq:msd_fitting}).
Both functions \(f_{\text{LE}}\) and \(f_{\text{mKV}}\) interpolate the asymptotic behaviors of (R)MSD in viscoelastic fluids. On short time scales, they correctly reproduce the ballistic behavior \( \sim t^2 \).
At long time scales, the behavior of the two functions shows different situations. \(f_{\text{LE}}\) approaches normal diffusion, while \(f_{\text{mKV}}\) converges to a constant value, reflecting the contribution of elasticity.
The distribution of the characteristic times \( \tau_j \) in \( H(\tau) \) determines the frequency window of the complex modulus \( G^*(\omega) \). 
In particular, the shortest timescale of \( H(\tau) \) determines the upper frequency range relevant to viscoelastic rheology that can be detected by (R)MSD.
By using fitting functions that incorporates the appropriate asymptotic behaviors of RMSD, we can effectively suppress the noise in the measured data while preserving the main features of viscoelastic rheology.
In the full RMR, 
 the nonlinear equation for \(\eta^{*}\)~[Eq.~(\ref{eq:rgle_mr})] was solved using the Newton--Raphson method.

\begin{figure}
    \centering
    \includegraphics[width=246.0pt]{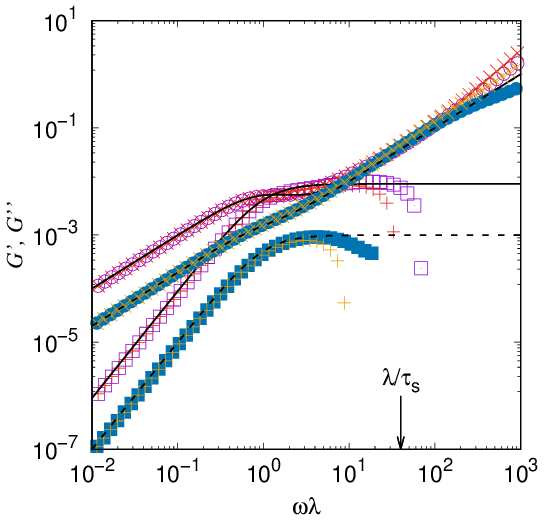}
    \caption{\label{fig:modulus_rmr}
    Comparison of the dynamic moduli derived using the inertialess RMR
    (red plus (\(G'\)) and X (\(G''\)) for \(\beta=0.1\), and yellow plus (\(G'\)) and X (\(G''\) for \(\beta=0.5\)) 
    and full RMR
    (purple open square (\(G'\)) and circle (\(G''\)) for \(\beta=0.1\), and blue closed square (\(G'\)) and circle (\(G''\)) for \(\beta=0.5\)) 
    with the input values of \(G'\) and \(G''\) (black lines).
    } 
\end{figure}

The dynamic moduli obtained by the inertialess RMR and the full RMR
are shown in Fig.~\ref{fig:modulus_rmr} and are compared with the input dynamic modulus.
The relaxation time from the RMR  is estimated by using the following equation: 
    \begin{align}
        \lambda_{\text{RMR}} &= \frac{
        \lim_{\omega\to 0}\frac{G''_{\text{RMR}}}{\omega}
        -
        \lim_{\omega\to \infty}\frac{G''_{\text{RMR}}}{\omega}
        }{\lim_{\omega\to\infty}G'_{\text{RMR}}},
    \end{align}
The estimated
relaxation times are \(\lambda_{\text{RMR}}\) = 982\(\tau_{0}\) and 951\(\tau_{0}\) for \(\beta=0.1\) and 0.5, respectively; they are reasonably
close to the input relaxation time \(\lambda\) = 1000\(\tau_{0}\).
As observed with the \(D_{r}\) values in RMSD,
the \(G'\) and \(G''\) obtained from the RMR  also show no system-size effects due to PBCs in all available frequencies, 
which is in contrast to the system size effects observed in translational MR~\cite{nakayama2024simulating,nakayama2025brownian}.

The \(G^{'}\) and \(G^{''}\) obtained by the inertialess RMR agree with the input ones in \(\omega\ll 1/\tau_{s}\), but the deviation becomes significant at high frequencies.
The loss modulus \(G^{''}\) obtained from the full RMR,  which takes into account inertial effects, shows an improved agreement with the input one up to \(\omega \approx 1/\tau_{s}\).
On the other hand, for the storage modulus \(G'\), the full RMR  provides slightly better accuracy at high frequencies compared to the inertialess RMR but is still underestimated around \(\omega \approx 1/\tau_{s}\).
This is likely due to the fact that at short times where \(\omega \agt 1/\tau_{s}\), \(G'\) is significantly smaller than \(G''\).
At this timescale, where inertial effects become significant, the RMSD is dominated by ballistic behavior and the elastic signal is relatively quite small, so it is difficult to quantitatively evaluate \(G'\) at \(\omega\agt 1/\tau_{s}\) even with the full RMR. %
In this high frequency region, the corresponding short-time RMSD is mainly dominated by ballistic motion, RMSD \( \sim t^2 \). In this region, the contribution of \( G' \) is relatively small compared to the ballistic behavior and, therefore, becomes almost undetectable from the RMSD. Since ballistic behavior itself implies \( G' \to 0 \), removing the inertial effect by the full RMR results in an obvious undershoot in \( G'_{\text{RMR}} \).
The results shown in Fig.~\ref{fig:modulus_rmr} demonstrate that the loss modulus \(G''\) can be quantitatively evaluated up to the vicinity of \(\omega \approx 1/\tau_{s}\) with full RMR. Furthermore, when \(G''/G'\) is small, the full RMR method can evaluate the storage modulus \(G'\) up to time scales several times larger than \(\tau_{s}\).

\section{\label{sec:conc}
Conclusions}
In this paper, passive rotational microrheology (RMR) was validated by evaluating the dynamic modulus \(G^*\) of a suspending fluid through the rotational Brownian motion of a spherical probe particle by using a direct numerical simulation of Brownian motion in a viscoelastic fluid.
The two methods of RMR 
were compared, specifically the inertialess RMR based on the Generalized Stokes--Einstein relation for rotational diffusion~(RGSER)~\cite{cheng2003rotational,andablo2005microrheology} and the full RMR  based on the generalized Langevin equation for rotation that takes into account fluid and particle inertia~\cite{zatovsky1969,berne1972hydrodynamic,Lisy2004}.
To analyze the rotational diffusion of a Brownian particle in a system with known fluid rheology, direct numerical simulations of the fluctuating Oldroyd-B fluid were performed using the stochastic Smoothed Profile method~\cite{nakayama2024simulating}.
The inertialess RMR is effective in estimating \(G^{*}\) for \(\omega\lambda \alt 1\), but the deviation from the true \(G^{*}\) becomes significant at high frequencies. This result is consistent with theoretical predictions based on rotational GLE.
In contrast, using the full RMR improves the accuracy of the \(G^{*}\) estimate up to the frequency \(\omega \approx \tau_{s}^{-1}=\eta_{s}/\rho_{f}a^{2}\), where fluid inertia becomes relevant.
On the other hand, in the ballistic regime of RMSD (\(t \ll \tau_{s}\)), 
where particle inertia is dominant, the contribution of fluid rheology is relatively small, making it difficult to accurately evaluate \(G^{*}\) even with the full RMR. %
This study has clarified the applicability range of the RMR, which utilizes the rotational Brownian motion of a probe particle.

Rotational microrheology is more suitable for localized analysis compared to translational microrheology. Therefore, it would be effective for the  rheological analysis of confined spaces, small sample volumes, and highly inhomogeneous media~\cite{yamada2000mechanics,yang2017application,liu2018rheological,xia2018microrheology,cai2021dynamic}.
Furthermore, it has been suggested that the analysis using rotational GLE can be extended to rheological properties utilizing dielectric relaxation~\cite{urakawa2023viscoelastic}.
Experimentally, measuring the rotational motion of a probe particle is more challenging than measuring translational motion. However, the measuring techniques for rotational motion of colloids have been reported using non-spherical particles~\cite{cheng2003rotational} and birefringent spherical particles~\cite{bishop2004optical,sandomirski2004highly,andablo2005microrheology,yanagishima2021particle}, making it possible to apply RMR using these methods.
Nevertheless, it should be noted that RMR is still sensitive to the quality of the rotational tracking which requires high spatial resolution for determining the central position of the probe particle and high temporal resolution for tracking the rotational motion.

Furthermore, rotational Brownian motion is insensitive to periodic boundary conditions and can, therefore, be directly applied to various types of mesoscale molecular simulations for evaluating the rheology of complex systems. This includes coarse-grained molecular dynamics, dissipative particle dynamics, and fluid dynamics simulations, as demonstrated in this study.
In conclusion, rotational microrheology offers a promising approach for detailed rheological analysis in various complex systems and conditions.

\begin{acknowledgments}
The numerical calculations were mainly carried 
out using the computer facilities at the Research Institute for 
Information Technology at Kyushu University and SQUID at D3 Center, Osaka University.
This work was 
supported by the Grants-in-Aid for Scientific Research (JSPS KAKENHI) 
under Grant No.~JP23K03343 
and the JSPS Core-to-Core Program ``Advanced core-to-core network 
for the physics of self-organizing active matter 
(No.~JPJSCCA20230002).'' 
Financial support from Hosokawa Powder Technology Foundation is also gratefully acknowledged.
\end{acknowledgments}

\section*{AUTHOR DECLARATIONS}
\subsection*{Conflict of Interest}
The author has no conflicts to disclose.

\subsection*{Author Contributions}
\noindent
Yasuya Nakayama: Conceptualization (equal); Data curation (equal); Formal analysis (equal); Funding acquisition (equal); Investigation (equal); Methodology (equal); Project administration (equal); Software (equal); Supervision (equal); Validation (equal); Visualization (equal); Writing - original draft (equal); Writing - review \& editing (equal).

\section*{Data Availability Statement}
The data that support the findings of this study are available from the corresponding author upon reasonable request.

\end{document}